\documentclass[journal, one column]{IEEEtran}
\usepackage{balance}
\usepackage{cite}
\usepackage{graphicx}
\usepackage{psfrag}
\usepackage{subfigure}
\usepackage{url}
\usepackage{stfloats}
\usepackage{amsmath}
\usepackage{array}
\usepackage{amssymb}
\usepackage{setspace}
\usepackage{algorithmicx}
\usepackage[ruled]{algorithm}
\usepackage{algpseudocode}
\usepackage{algpascal}
\usepackage{algc}
\usepackage{setspace}

\begin{document}

\markboth{Submitted to IEEE Transactions on Vehicular Technology, as Correspondence}%
{Shell \MakeLowercase{\textit{et al.}}: Bare Demo of IEEEtran.cls
for Journals}

\title{Use of Non-Orthogonal Multiple Access in Dual-hop relaying}
\author{Wei Duan, Miaowen Wen,~\IEEEmembership{Member,~IEEE,} Yier Yan,~Zixiang Xiong,~\IEEEmembership{Fellow, IEEE,}\\~and Moon Ho
Lee,~\IEEEmembership{Senior Member, IEEE}
\thanks{W. Duan and M. H. Lee are with the Institute of Information and Communication, Chonbuk National University, 664-14 Deokjin-dong, Jeonju 561-756, Republic of Korea. (e-mail:
sinder@live.cn, moonho@jbnu.ac.kr)}
\thanks{M. Wen is with School of Electronic and Information Engineering,
South China University of Technology, Guangzhou 510640, China.
(e-mail: eemwwen@scut.edu.cn.)}
\thanks{Y. Yan is with School of Mechanical and Electrical Engineering,
Guangzhou University, Guangzhou, China. (e-mail:
year0080@gzhu.edu.cn.)}
\thanks{Z. Xiong is with the Department of Electrical Engineering, Texas A $\&$ M University, College Station, TX 77843 USA. (email: zx@ece.tamu.edu.)}
} \maketitle

\begin{abstract}

\begin{spacing}{2.0}
To improve the sum-rate (SR) of the dual-hop relay system, a novel
two-stage power allocation scheme with non-orthogonal multiple
access (NOMA) is proposed. In this scheme, after the reception of
the superposition coded symbol with a power allocation from the
source, the relay node forwards a new superposition coded symbol
with another power allocation to the destination. By employing the
maximum ratio combination (MRC), the destination jointly decodes
the information symbols from the source and the relay. Assuming
Rayleigh fading channels, closed-form solution of the ergodic SR
at high signal-to-noise ratio (SNR) is derived and a practical
power allocation is also designed for the proposed NOMA scheme.
Through numerical results, it is shown that the performance of the
proposed scheme is significantly improved compared with the
existing work. \end{spacing}

\end{abstract}
\\

\section{Introduction}
\begin{spacing}{2.0}
Recently, non-orthogonal multiple access (NOMA) technique is
widely considered as a promising multiple access (MA) candidate
for 5G mobile networks due to its superior spectral efficiency
\cite{a1}, \cite{a2}. The key idea of NOMA is to explore the power
domain for realizing MA, where different users are served at
different power levels \cite{5} and \cite{1}. Consequently, the
impact of different choices of power allocation coefficients has
been studied in \cite{a5}, where the authors considered the
fairness of the user for NOMA systems. Under the total transmit
power constraint and the minimum rate constraint of the weak user,
the ergodic capacity maximization problem for multiple-input
multiple-output (MIMO) NOMA systems has been solved \cite{a6}.

In addition, the implementation of NOMA with two base stations
(BSs) has been investigated in \cite{a4}. The authors derived the
sum rates (SRs) for NOMA schemes with superposition coding (SC)
that does not require instantaneous channel state information
(CSI) at BSs. On this basis, the work of NOMA in the coordinated
direct and relay transmission (CDRT) has been introduced in
\cite{1}. The authors proposed the cooperative relaying system
(CRS) using NOMA, and presented the exact and asymptotic
expressions for the achievable rate of the proposed system in
independent Rayleigh fading channels. Nevertheless, the achievable
rate is limited by the source-to-destination link which leading to
the loss of the system performance.

To solve this problem, we propose a two-stage power allocation CRS
using NOMA in this paper. In our proposed scheme, once MRC at the
destination and twice power allocations at the source and the
relay are employed. To further improve the performance of the
ergodic SR of our proposed scheme and the existing work \cite{1},
the destination will not immediately decode the reception from the
source until it receives the superposition coded signal from the
relay. We focus on the analysis of the achievable ergodic SR for
our proposed scheme, and the closed-form expression at SNR region
is then derived. In addition, to maximize the ergodic SR, a
practical two-stage power allocation strategy is designed.
Numerical results are presented to corroborate the derived
theoretical analysis, and show that the proposed scheme
significantly improves the ergodic SR compared to the one in
\cite{1}.

\section{System model and proposed scheme}

\begin{figure}
\begin{center}
\includegraphics [width=70mm]{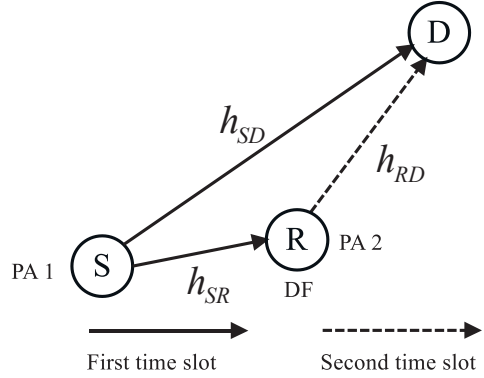}\\
\caption{\label{9} The Dual-hop relay system with two-stage power
allocation.}
\end{center}
\end{figure}

A simple CRS consisting of one source, one relay, and one
destination is shown in Fig. 1. We assume all nodes operate in
half-duplex mode and the direct link between the source and the
destination exists. The channels from the source to the
destination, from the source to the relay, and from the relay to
the destination are denoted as $h_{SD}$, $h_{SR}$, and $h_{RD}$,
respectively, which are assumed to be independent complex Gaussian
random variables with variances $\alpha_{SD}$, $\alpha_{SR}$, and
$\alpha_{RD}$, respectively. In our proposed scheme, each
transmission involves two time slots. At the first time slot,
assuming the adoption of the superposition code, the signal
$\sqrt{a_{1}P_{t}}x_{1}+\sqrt{a_{2}P_{t}}x_{2}$ is simultaneously
transmitted from the source to the relay and the destination,
where $x_{i}$, i=1,2, denote the broadcasted symbols at the
source, $a_{1}$ and $a_{2}$ with $a_{1}+a_{2}=1$ are the power
allocation factors, and $P_{t}$ stands for the total transmit
power. The received signal at the source and the relay are given
by
\begin{eqnarray}\label{1}
y_{R}&=&h_{SR}\left(\sqrt{a_{1}P_{t}}x_{1}+\sqrt{a_{2}P_{t}}x_{2}\right)+n_{R},\\
y^{(1)}_{D}&=&h_{SD}\left(\sqrt{a_{1}P_{t}}x_{1}+\sqrt{a_{2}P_{t}}x_{2}\right)+n^{(1)}_{D},
\end{eqnarray}
where $\left\{n_{R},n^{(1)}_{D}\right\}\sim{CN(0,\sigma^{2})}$
represent the additive white Gaussian noises (AWGNs) with zero
mean and variance $\sigma^{2}$.

To successfully and simultaneously decode $x_{1}$ and $x_{2}$ at
the relay, we further assume that the path loss and shadowing
effects for $h_{SD}$ are worse than those for $h_{SR}$, which
leading the ordinary channel condition
as $\alpha_{SD}<\alpha_{SR}$. 
By treating $x_{2}$ as noise to decode $x_{1}$, and then using the
SIC to acquire $x_{2}$, the relay can efficiently decode $x_{1}$
and $x_{2}$ from \eqref{1} during one time slot. By this way, the
SNRs for $x_1$ and $x_2$ at the relay can be respectively
expressed as
\begin{eqnarray}\label{2}
\gamma_{R}^{\left(x_{1}\right)}=\frac{\left|h_{SR}\right|^{2}a_{1}P_{t}}{\left|h_{SR}\right|^{2}a_{2}P_{t}+\sigma^{2}}
~,~
\gamma_{R}^{\left(x_{2}\right)}=\frac{\left|h_{SR}\right|^{2}a_{2}P_{t}}{\sigma^{2}}\label{a2}.
\end{eqnarray}
Note that different from \cite{1}, to improve the system
performance, the destination will not decode the received signal
from the source but instead conserves it until the second time
slot comes.


At the second time slot, the relay node forwards a new symbol
$x_{R}$ with SC to the destination:
\begin{eqnarray}\label{3}
x_{R}=\sqrt{a_{3}P_{t}}x_{1}-\sqrt{a_{4}P_{t}}x_{2},
\end{eqnarray}
where $P_{t}$ is the total transmit power, $a_{3}$ and $a_{4}$
with $a_{3}+a_{4}=1$ are new power allocation coefficients.
Therefore, the received signal at the destination can be written
as
\begin{eqnarray}\label{4}
y_{D}^{(2)}=h_{RD}\left(\sqrt{a_{3}P_{t}}x_{1}-\sqrt{a_{4}P_{t}}x_{2}\right)+n_{D}^{(2)},
\end{eqnarray}
where $n_{D}^{(2)}$ is the AWGN at the destination with zero mean
and variance $\sigma^{2}$. From above, we see that there are two
signals received at the destination, which are $y_{D}^{(1)}$ and
$y_{D}^{(2)}$. To jointly decode $x_{1}$ and $x_{2}$ at the
destination, we employ
$y_{D}^{(1)}\sqrt{a_{4}}h_{R,D}+y_{D}^{(2)}\sqrt{a_{2}}h_{S,D}$
and
$y_{D}^{(1)}\sqrt{a_{4}}h_{R,D}-y_{D}^{(2)}\sqrt{a_{2}}h_{S,D}$.
Therefore, the target signals can be expressed as
\begin{eqnarray}\label{5}
\!\!\!T_{x_{1}}\!\!\!\!\!&=&\!\!\!\!\!h_{SD}h_{RD}\varsigma\sqrt{P_{t}}x_{1}+\sqrt{a_{4}}h_{RD}n_{D}^{(1)}+\sqrt{a_{2}}h_{SD}n_{D}^{(2)},\\
\!\!\!T_{x_{2}}\!\!\!\!\!&=&\!\!\!\!\!h_{SD}h_{RD}\varsigma\sqrt{P_{t}}x_{2}+\sqrt{a_{3}}h_{RD}n_{D}^{(1)}+\sqrt{a_{1}}h_{SD}n_{D}^{(2)},\label{7}
\end{eqnarray}
where $\varsigma=\sqrt{a_{1}a_{4}}+\sqrt{a_{2}a_{3}}$. According
to \eqref{5} and \eqref{7}, the corresponding receive SNR for
$x_{1}$ and $x_{2}$ can be obtained as
\begin{eqnarray}\label{6}
\gamma_{D}^{\left(x_{1}\right)}=\frac{\left|h_{SD}\right|^{2}\left|h_{RD}\right|^{2}\varsigma^{2}\rho}{a_{4}\left|h_{RD}\right|^{2}+a_{2}\left|h_{SD}\right|^{2}},
\end{eqnarray}
and
\begin{eqnarray}\label{8}
\gamma_{D}^{\left(x_{2}\right)}=\frac{\left|h_{SD}\right|^{2}\left|h_{RD}\right|^{2}\varsigma^{2}\rho}{a_{3}\left|h_{RD}\right|^{2}+a_{1}\left|h_{SD}\right|^{2}},
\end{eqnarray}
respectively, where $\rho=\frac{P_{t}}{\sigma^{2}}$ denotes the
transmit SNR.

\section{Achievable SR and performance analysis}

In this section, in order to characterize the superiority of our
proposed scheme, we will focus on the analysis the achievable
ergodic SR and derive its closed-form expression at high SNR.

The achievable rate associated with symbol $x_{1}$ is obtained
using \eqref{2} and \eqref{6} as
\begin{eqnarray}\label{q1}
C_{x_{1}}&=&\mathrm{min}\left\{\frac{1}{2}\mathrm{log}\left(1+\gamma_{D}^{(x_{1})}\right),\frac{1}{2}\mathrm{log}\left(1+\gamma_{R}^{(x_{1})}\right)\right\}\notag\\
&=&\frac{1}{2}~\mathrm{log}\left(1+\mathrm{min}\left\{\gamma_{D}^{(x_{1})},\gamma_{R}^{(x_{1})}\right\}\right),
\end{eqnarray}
where $1/2$ is resulted from the dual-hop transmission in two time
slots. Similarly, according to \eqref{a2} and \eqref{8}, the
achievable rate associated with symbol $x_{2}$ is obtained as
\begin{eqnarray}\label{q2}
C_{x_{2}}&=&\mathrm{min}\left\{\frac{1}{2}\mathrm{log}\left(1+\gamma_{D}^{(x_{2})}\right),\frac{1}{2}\mathrm{log}\left(1+\gamma_{R}^{(x_{2})}\right)\right\}\notag\\
&=&\frac{1}{2}~\mathrm{log}\left(1+\mathrm{min}\left\{\gamma_{D}^{(x_{2})},\gamma_{R}^{(x_{2})}\right\}\right)\notag\\
&=&\frac{1}{2}~\mathrm{log}\left(1+\mathrm{min}\left\{\widetilde{\gamma}_{D}^{(x_{2})},\widetilde{\gamma}_{R}^{(x_{2})}\right\}\rho\right),
\end{eqnarray}
where
$\widetilde{\gamma}_{D}^{(x_{i})}=\frac{1}{\rho}\gamma_{D}^{(x_{i})}$,
for $i=1,2$. Synthesizing \eqref{q1} and \eqref{q2}, the
achievable SR can be expressed as
\begin{eqnarray}\label{qq1}
C_{sum}&=&C_{x_{1}}+C_{x_{2}}.
\end{eqnarray}
Further denoting $\left|h_{SR}\right|^{2}=\beta_{SR}$,
$\left|h_{RD}\right|^{2}=\beta_{RD}$, and
$\left|h_{SD}\right|^{2}=\beta_{SD}$, we have
\begin{eqnarray}\label{10}
\mathrm{min}\{\gamma_{D}^{\left(x_{1}\right)},\gamma_{R}^{(x_{1})}\}
=\mathrm{min}\left\{\frac{\varsigma^{2}\rho
\beta_{SD}\beta_{RD}}{a_{2}\beta_{SD}+a_{4}\beta_{RD}},
\frac{\beta_{SR}a_{1}\rho}{\beta_{SR}a_{2}\rho+1}\right\},\notag
\end{eqnarray}
and
\begin{eqnarray}\label{11}
\mathrm{min}\{\widetilde{\gamma}_{D}^{(x_{2})},\widetilde{\gamma}_{R}^{(x_{2})}\}=\mathrm{min}\left\{\frac{\varsigma^{2}\beta_{SD}
\beta_{RD}}{a_{1}\beta_{SD}+a_{3}\beta_{RD}},
a_{2}\beta_{SR}\right\}.\notag
\end{eqnarray}
Letting
$X=\mathrm{min}\{\gamma_{D}^{(x_{1})},\gamma_{R}^{(x_{1})}\}$, the
complementary cumulative distribution function (CCDF) of $X$ can
be obtained as
\begin{eqnarray}\label{e1}
\overline{F}_{X}(x)\!=\!Pr\!\left\{\frac{m\rho
\beta_{SD}\beta_{RD}}{a_{2}\beta_{SD}+a_{4}\beta_{RD}}\!>\!x,
\frac{\beta_{SR}a_{1}\rho}{\beta_{SR}a_{2}\rho+1}\!>\!x\right\}.
\end{eqnarray}
Noting that the CCDF of
$\beta_{\delta}=e^{-\frac{x}{\alpha_{\delta}}}$, for
$\delta\in\{SR, SD, RD\}$, \eqref{e1} can be equivalently
represented as
\begin{eqnarray}\label{e2}
\overline{F}_{X}(x)&=&\overline{F}_{SR}\left(\frac{x}{a_{1}\rho-a_{2}\rho
x}\right)\Bigg[Pr\left\{\beta_{SD}>\frac{a_{4}\beta_{RD}x}{\varsigma^{2}\rho\beta_{RD}-a_{2}x}\bigg\vert~\beta_{RD}>\frac{a_{2}x}{\varsigma^{2}\rho}\right\}\overline{F}_{RD}\left(\frac{a_{2}x}{\varsigma^{2}\rho}\right)\notag\\
&&+Pr\left\{\beta_{SD}<\frac{a_{4}\beta_{RD}x}{\varsigma^{2}\rho\beta_{RD}-a_{2}x}\bigg\vert~\beta_{RD}<\frac{a_{2}x}{\varsigma^{2}\rho}\right\}
F_{RD}\left(\frac{a_{2}x}{\varsigma^{2}\rho}\right)\Bigg]\notag\\
&=&\frac{1}{\alpha_{RD}}e^{-\frac{x}{\left(a_{1}\rho-a_{2}\rho
x\right)\alpha_{SR}}-\frac{a_{2}x}{\varsigma^{2}\rho\alpha_{RD}}}\int^{\infty}_{\frac{a_{2}x}{\varsigma^{2}\rho}}e^{-\frac{a_{4}u
x}{\left(\varsigma^{2}\rho
u-a_{2}x\right)\alpha_{SD}}-\frac{u}{\alpha_{RD}}}du\notag\\
&&+\frac{1}{\alpha_{RD}}e^{-\frac{x}{\left(a_{1}\rho-a_{2}\rho
x\right)\alpha_{SR}}}\times\left(1-e^{-\frac{a_{2}x}{\varsigma^{2}\rho\alpha_{RD}}}\right)\int^{\frac{a_{2}x}{\varsigma^{2}\rho}}_{0}e^{-\frac{a_{4}u
x}{\left(\varsigma^{2}\rho
u-a_{2}x\right)\alpha_{SD}}-\frac{u}{\alpha_{RD}}}du.
\end{eqnarray}
Consider the high transmit SNR case, i.e., $\rho\gg 1$. In this
case, we have
\begin{eqnarray}\label{m1}
\frac{\beta_{SR}a_{1}\rho}{\beta_{SR}a_{2}\rho+1}\sim\frac{a_{1}}{a_{2}}.
\end{eqnarray}
Letting $t_{1}=\varsigma^{2}\rho u-a_{2}x$ and using
$\int_{0}^{\infty}e^{-\frac{a}{x}-bx}dx=2\sqrt{\frac{a}{b}}K_{1}\left(2\sqrt{ab}\right)$
\cite[3.324.1]{3}, when $x<\frac{a_{1}}{a_{2}}$,
\eqref{e2} can be equivalently written as
\begin{eqnarray}\label{e5}
\overline{F}_{X}(x)&=&\frac{e^{-\frac{a_{2}x}{\varsigma^{2}\rho\alpha_{RD}}-\frac{a_{4}x}{\varsigma^{2}\rho\alpha_{SD}}}}{{\varsigma^{2}\rho\alpha_{RD}}}\int^{\infty}_{0}e^{-\frac{a_{2}a_{4}
x^{2}}{\varsigma^{2}\rho
t_{1}\alpha_{SD}}-\frac{t_{1}}{\varsigma^{2}\rho\alpha_{RD}}}dt_{1}\notag\\
&=&\frac{2x}{{\varsigma^{2}\rho\xi\phi_{1}\alpha_{RD}}}e^{-\frac{a_{2}x}{\varsigma^{2}\rho\alpha_{RD}}-\frac{a_{4}x}{\varsigma^{2}\rho\alpha_{SD}}}K_{1}\!\!\left(\frac{2\xi
x}{\rho\phi_{1}}\right),
\end{eqnarray}
where $\xi=\sqrt{\frac{1}{\varsigma^{2}\alpha_{RD}}}$,
$\phi_{1}=\sqrt{\frac{\varsigma^{2}\alpha_{SD}}{a_{2}a_{4}}}$, and
$K_{1}(\cdot)$, denotes the first order modified Bessel function
of the second kind \cite{2}. For the case $x>\frac{a_{1}}{a_{2}}$,
$\overline{F}_{X}(x)=0$ always holds true due to
$\frac{\beta_{SR}a_{1}\rho}{\beta_{SR}a_{2}\rho+1}<\frac{a_{1}}{a_{2}}$.
The achievable rate can be calculated as
\begin{eqnarray}\label{m2}
\widetilde{C}_{x_{1}}&=&
\int_{0}^{\frac{a_{1}}{a_{2}}}\frac{1}{2}\mathrm{log}_{2}\left(1+x\right)dF_{X}(x)
+\frac{1}{2}\mathrm{log}_{2}\left(1+\frac{a_{1}}{a_{2}}\right)\left(1-F_{X}\left(\frac{a_{1}}{a_{2}}\right)\right)\notag\\
&=&\frac{1}{2}\mathrm{log}_{2}\left(1+\frac{a_{1}}{a_{2}}\right)-\frac{1}{2\mathrm{ln}2}\int_{0}^{\frac{a_{1}}{a_{2}}}\frac{1}{1+x}\left(1-\frac{1}{{m\rho\xi\phi_{x}\alpha_{RD}}}2xe^{-\frac{a_{2}x}{m\rho\alpha_{RD}}-\frac{a_{4}x}{m\rho\alpha_{SD}}}K_{1}\left(\frac{2\xi
x}{\rho\phi_{x}}\right)\right)dx,
\end{eqnarray}
where the first equality holds due to $\int_0^\infty
{\frac{1}{2}{{\log}_2}\left( {1 + x}\right){f_{X}}}
\left(x\right)dx=\frac{1}{{2\ln 2}}\int_0^\infty {\frac{{1-
{F_{X}}\left(x\right)}}{{1+x}}}dx$ with
$F_{X}(x)=1-\overline{F}_{X}(x)$.
For small $x$,
$K_{\nu}(x)\approx\frac{\Gamma(\nu)}{2}\left(\frac{2}{x}\right)^{\nu}$
\cite{d1}, where $\Gamma(\cdot)$ denotes the gamma function.
Therefore, \eqref{m2} can be approximately rewritten as
\begin{eqnarray}\label{v1}
\widetilde{C}_{x_{1}}&=&\frac{1}{2}\mathrm{log}_{2}\left(1+\frac{a_{1}}{a_{2}}\right)\!\!-\!\!\frac{1}{{2\mathrm{ln}2}}\int_{0}^{\frac{a_{1}}{a_{2}}}\frac{1\!-\!e^{-\frac{a_{2}x}{\varsigma^{2}\rho\alpha_{RD}}-\frac{a_{4}x}{\varsigma^{2}\rho\alpha_{SD}}}}{1+x}dx\notag\\
&=&\frac{e^{\frac{1}{\varsigma^{2}\rho}\big(\frac{a_{2}}{\alpha_{RD}}+\frac{a_{4}}{\alpha_{SD}}\big)}}{2\mathrm{ln}2}\Bigg[\mathrm{Ei}\left(-\frac{a_{1}}{a_{2}\varsigma^{2}\rho}\left(\frac{a_{2}}{\alpha_{RD}}+\frac{a_{4}}{\alpha_{SD}}\right)\right)-\mathrm{Ei}\left(-\frac{1}{\varsigma^{2}\rho}\left(\frac{a_{2}}{\alpha_{RD}}+\frac{a_{4}}{\alpha_{SD}}\right)\right)\Bigg],
\end{eqnarray}
where $\mathrm{Ei}\left(\cdot\right)$ denotes the exponential
integral function, and the integral result
$\int_{0}^{u}\frac{e^{-\mu
x}dx}{x+\beta}=e^{\mu\beta}\left[\mathrm{Ei}\left(-\mu
u-\mu\beta\right)-\mathrm{Ei}\left(-\mu\beta\right)\right]$
\cite[3.352.1]{3} is used in the first equality. Considering high
SNR case, \eqref{10} can be equivalently rewritten as
\begin{eqnarray}\label{10a}
\mathrm{min}\{\gamma_{D}^{\left(x_{1}\right)},\gamma_{R}^{(x_{1})}\}=\mathrm{min}\left\{\rho\times\frac{m
\beta_{SD}\beta_{RD}}{a_{2}\beta_{SD}+a_{4}\beta_{RD}},
\frac{a_{1}}{a_{2}}\right\}\triangleq\frac{a_{1}}{a_{2}},
\end{eqnarray}
which leading
$C_{x_{1}}\triangleq\frac{1}{2}\mathrm{log}_{2}\left(1+\frac{a_{1}}{a_{2}}\right)$.
Also, for \eqref{v1}, we have
$\frac{1}{\varsigma^{2}\rho}\left(\frac{a_{2}}{\alpha_{RD}}+\frac{a_{4}}{\alpha_{SD}}\right)\approx0$
with $\rho\gg1$ and the achievable rate
$\widetilde{C}_{x_{1}}\triangleq\frac{1}{2}\mathrm{log}_{2}\left(1+\frac{a_{1}}{a_{2}}\right)$.
It is clear that there is an good match between the analytical
result in \eqref{v1} and \eqref{10}.
Similarly, letting
$Y=\mathrm{min}\{\widetilde{\gamma}_{D}^{(x_{2})},\widetilde{\gamma}_{R}^{(x_{2})}\}$
the CCDF of $Y$ can be obtained as
\begin{eqnarray}\label{ee1}
\overline{F}_{Y}(y)=\overline{F}^{(1)}_{Y}(y)\times\overline{F}^{(2)}_{Y}(y),
\end{eqnarray}
where $\overline{F}^{(1)}_{Y}(y)=Pr\left\{\frac{\varsigma^{2}
\beta_{SD}\beta_{RD}}{a_{1}\beta_{SD}+a_{3}\beta_{RD}}>y\right\}$
and
$\overline{F}^{(2)}_{Y}(y)=Pr\left\{a_{2}\beta_{SR}>y\right\}$.
Since $\overline{F}^{(1)}_{Y}(y)$ can be rewritten as
\begin{eqnarray}\label{ww1}
\overline{F}^{(1)}_{Y}(y)=Pr\left\{\frac{\varsigma^{2}}{2a_{1}a_{3}}\cdot\frac{2
a_{1}a_{3}\beta_{SD}\beta_{RD}}{a_{1}\beta_{SD}+a_{3}\beta_{RD}}>y\right\},
\end{eqnarray}
using \cite[Theorem 1]{4}, we have
\begin{eqnarray}\label{ww2}
\overline{F}^{(1)}_{Y}(y)=\frac{2y}{\varsigma^{2}}\sqrt{\frac{a_{1}a_{3}}{\alpha_{SD}\alpha_{RD}}}
e^{-\frac{\left(a_{1}\alpha_{SD}+a_{3}\alpha_{RD}\right)y}{\varsigma^{2}\alpha_{SD}\alpha_{RD}}}K_{1}\left(\frac{2y}{\varsigma^{2}}\sqrt{\frac{a_{1}a_{3}}{\alpha_{SD}\alpha_{RD}}}
\right).
\end{eqnarray}
Therefore, the CCDF of $Y$ can be finally expressed as
\begin{eqnarray}\label{ww3}
\overline{F}_{Y}(y)=\frac{2\gamma
y}{\rho}e^{-\big(\frac{a_{1}\alpha_{SD}+a_{3}\alpha_{RD}}{\varsigma^{2}\alpha_{SD}\alpha_{RD}}+\frac{1}{a_{2}\alpha_{SR}}\big)\frac{y}{\rho}}K_{1}\left(\frac{2\gamma
y}{\rho}\right),
\end{eqnarray}
where
$\gamma=\frac{1}{\varsigma^{2}}\sqrt{\frac{a_{1}a_{3}}{\alpha_{SD}\alpha_{RD}}}$.
As a double check, the CDF $F_{Y}(y)=1-\overline{F}_{Y}(y)$ in
\eqref{ww3} was validated by Monte Carlo simulation, as
illustrated in Fig. 2. It is clear that there is an excellent
match between the analytical result in \eqref{ww3} and the Monte
Carlo simulation.

\begin{figure}
\begin{center}
\includegraphics [width=90mm,height=75mm]{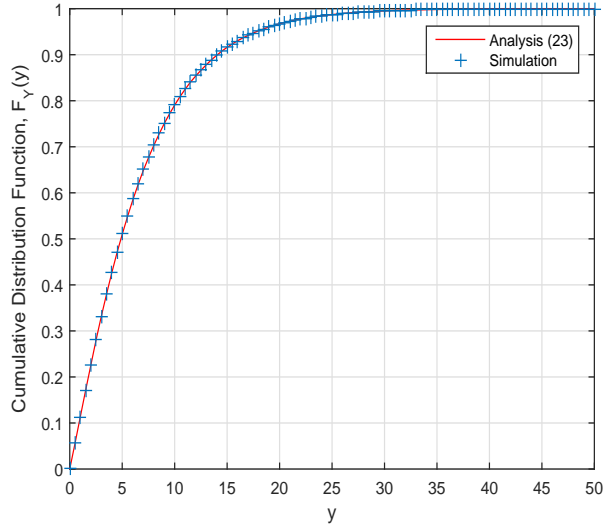}\\
\caption{\label{9} Comparison between the analytical results for
the CDF \eqref{ww3} and the Monte Carlo simulation.}
\end{center}
\end{figure}
Assuming
$\left(\frac{a_{1}\alpha_{SD}+a_{3}\alpha_{RD}}{\varsigma^{2}\alpha_{SD}\alpha_{RD}}+\frac{1}{a_{2}\alpha_{SR}}\right)\frac{1}{\rho}=\eta$,
and using \cite[3.352.2]{3}, the achievable rate of $x_{2}$ can be
obtained as
\begin{eqnarray}\label{yy1}
\widetilde{C}_{x_{2}}=\frac{\gamma}{\rho\mathrm{ln}2}\int_{0}^{\infty}\frac{ye^{-\eta
y}K_{1}\left( \frac{2\gamma
y}{\rho}\right)}{1+y}dy=-\frac{e^{\eta}}{2\mathrm{ln}2}\mathrm{Ei}\left(-\eta\right).
\end{eqnarray}
Finally, putting \eqref{v1} and \eqref{yy1} together, we can
express the ergodic SR of our proposed system in closed form as
\begin{eqnarray}\label{yy2}
\widehat{C}_{sum}=\frac{e^{\frac{1}{\varsigma^{2}\rho}\big(\frac{a_{2}}{\alpha_{RD}}+\frac{a_{4}}{\alpha_{SD}}\big)}}{2\mathrm{ln}2}\Bigg[\mathrm{Ei}\left(-\frac{a_{1}}{a_{2}\varsigma^{2}\rho}\left(\frac{a_{2}}{\alpha_{RD}}+\frac{a_{4}}{\alpha_{SD}}\right)\right)-\mathrm{Ei}\!\left(\!-\frac{1}{\varsigma^{2}\rho}\!\left(\frac{a_{2}}{\alpha_{RD}}\!+\!\frac{a_{4}}{\alpha_{SD}}\!\right)\!\right)\!\Bigg]-\frac{e^{\eta}}{2\mathrm{ln}2}\mathrm{Ei}\left(-\eta\right).
\end{eqnarray}



\section{Power Allocations}

This section presents appropriate power allocations of the source
and the relay for our proposed scheme. Using the approximations of
$\mathrm{Ei}(-x)=Ec+\mathrm{ln}(x)$ and $e^{x}=1+x$, for small
$x$, where $Ec$ denotes the Euler constant, the ergodic SR in
\eqref{yy2} can be approximately expressed as
\begin{eqnarray}\label{yy3}
\widetilde{C}_{sum}\!\!\!\!\!\!&\sim&\!\!\!\!\!\!\frac{1}{2}\!\mathrm{log}_{2}\!\left(\frac{a_{1}}{a_{2}}\right)\!\!\!\left(1\!\!+\!\!\frac{1}{\varsigma^{2}\rho}\!\!\left(\frac{a_{2}}{\alpha_{RD}}\!+\!\frac{a_{4}}{\alpha_{SD}}\right)\right)\!-\!\frac{1+\left(\frac{a_{1}\alpha_{SD}+a_{3}\alpha_{RD}}{\varsigma^{2}\alpha_{SD}\alpha_{RD}}+\frac{1}{a_{2}\alpha_{SR}}\right)\frac{1}{\rho}}{2\mathrm{ln}2}\left(\!Ec\!+\!\mathrm{ln}\!\left(\!\left(\frac{a_{1}\alpha_{SD}\!+\!a_{3}\alpha_{RD}}{\varsigma^{2}\alpha_{SD}\alpha_{RD}}\!+\!\frac{1}{a_{2}\alpha_{SR}}\right)\!\!\frac{1}{\rho}\right)\right)\notag\\
\!\!\!\!\!\!&\sim&\!\!\!\!\!\!\frac{1}{2}\mathrm{log}_{2}\left(\frac{a_{1}\varsigma^{2}\alpha_{SD}\alpha_{RD}\alpha_{SR}}{a_{2}\alpha_{SR}(a_{1}\alpha_{SD}+a_{3}\alpha_{RD})+\varsigma^{2}\alpha_{SD}\alpha_{RD}}\right)-\frac{Ec}{2\mathrm{ln}2}+\frac{1}{2}\mathrm{log}_{2}\rho.
\end{eqnarray}
Letting the derivative of \eqref{yy3} with respect to the power
allocation factors $a_{1}$ and $a_{3}$ be $0$, i.e.,
$\frac{\partial \widetilde{R}_{sum}}{\partial a_{i}}=0$, for
$i=1,3$, and assuming
$\varphi_{1}=\sqrt{\frac{1-a_{3}}{a_{1}}}-\sqrt{\frac{a_{3}}{1-a_{1}}}$,
$\varphi_{2}=\sqrt{\frac{a_{1}}{1-a_{3}}}-\sqrt{\frac{1-a_{1}}{a_{3}}}$,
and $\Psi=\sqrt{a_{1}(1-a_{3})}+\sqrt{(1-a_{1})a_{3}}$, the
optimal power allocation factor $a_{1}$ and $a_{2}$, for
$0<a_{i}<1$, can be obtained from \eqref{yy5} and \eqref{yy6}.
\begin{eqnarray}\label{yy5}
\frac{a_{1}\Psi\left(\varphi_{1}\Psi\alpha_{RD}\alpha_{SD}+\alpha_{SR}\alpha_{SD}-a_{3}\alpha_{SR}\alpha_{RD}-2a_{1}\alpha_{SD}\alpha_{SR}\right)}{
\Psi^2\alpha_{RD}\alpha_{SD}+ (1 - a_{1}) (a_{3}\alpha_{RD}+a_{1}\alpha_{SD})\alpha_{SR}}-a_{1}\varphi_{1}-\Psi&=&0,\\
\frac{a_{1}\Psi\left(\varphi_{2}\Psi\alpha_{RD}\alpha_{SD}+(1-a_{1})\alpha_{RD}\alpha_{SR}\right)}{
\Psi^2\alpha_{RD}\alpha_{SD}+ (1 - a_{1})
(a_{3}\alpha_{RD}+a_{1}\alpha_{SD})\alpha_{SR}}-a_{1}\varphi_{2}&=&0.\label{yy6}
\end{eqnarray}

\section{Numerical Results}

In this section, we examine the performance of our proposed
two-stage power allocation NOMA for CRN in terms of the ergodic SR
with fixed $\alpha_{SD}=1$. All results are averaged over $20,000$
channel realizations. Comparisons are made with the CRS-NOMA
\cite{1} in two considered system setups: (1) $\alpha_{SR}=10$,
$\alpha_{RD}=2$; (2): $\alpha_{SR}=2$, $\alpha_{RD}=10$.

\begin{figure}
\begin{center}
\includegraphics [width=90mm,height=75mm]{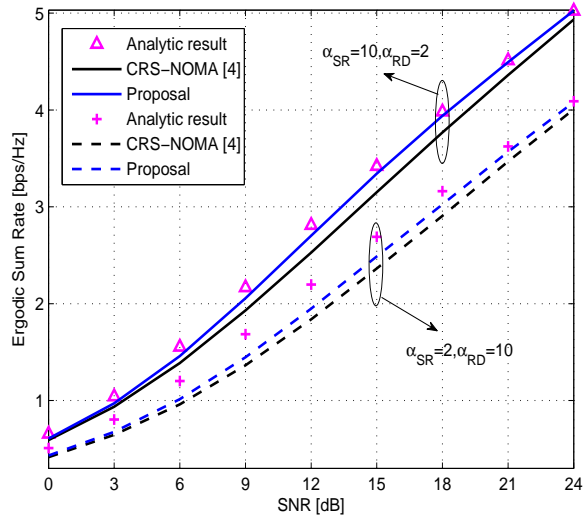}\\
\caption{\label{9} The ergodic SRs achieved by our proposal and
CRS-NOMA \cite{1} versus the transmit SNR.}
\end{center}
\end{figure}

Fig. 3 depicts the ergodic SR performance versus the transmit SNR
with fixed $a_{1}=0.95$ and $a_{3}=0.05$. It is easy to see that
there is a good match between the analytical result in \eqref{yy2}
and the simulation result. Furthermore, the ergodic SR of our
proposal outperforms the one in \cite{1}. This is because the
achievable rate is dominated by the receive SNR at the destination
which is always smaller than $C_{x_{1}}$. In addition, by
employing the two-stage power allocation and the MRC, the SNR
gains can be further improved. Remarkably, there is a gap between
the simulation results and the analytic results in for the small
$\alpha_{SR}$, this is because that the approximations of
$\frac{\alpha_{SR}a_{1}\rho  }{}$

\begin{figure}
\begin{center}
\includegraphics [width=90mm,height=75mm]{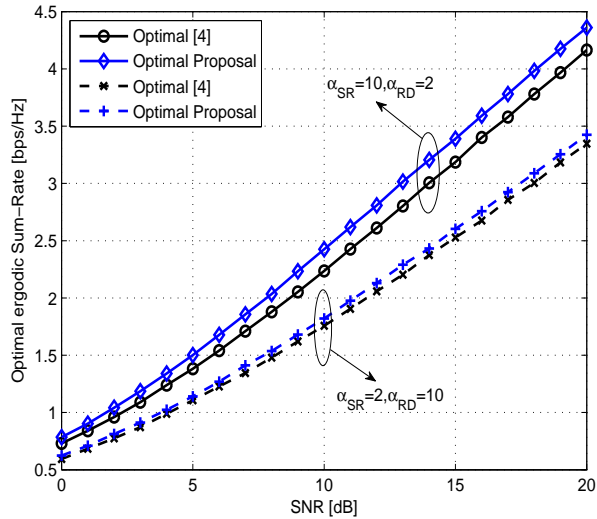}\\
\caption{\label{9} The ergodic SRs achieved by our proposed scheme
and CRS-NOMA \cite{1} with the optimal allocation schemes versus
the transmit SNR.}
\end{center}
\end{figure}

By using the exhaustive search to obtain the optimal power
allocation factor $\left\{a_{1}, a_{3}\right\}$ and
\eqref{yy3}-\eqref{yy6} to obtain the suboptimal one, in Fig. 4
and Fig. 5 we compare the optimal and sub optimal ergodic SR for
our proposed scheme versus the transmit SNR with the optimal one
in \cite{1}. Fig. 4 shows that the optimal ergodic SR of our
proposal overwhelms the optimal one for CRS-NOMA in the $0\sim20$
$\mathrm{dB}$ SNR region. In Fig. 5, it easy to see that the
suboptimal solution is close to the optimal one at high SNR, which
supports the practical utility of our design. Remarkably, in the
typically high SNR region, the performance of our proposed scheme
and CRS-NOMA are close. This observation is consistent with
\eqref{yy3} and \cite[Eq. (15)]{1}, where these equations are
resulted as $\frac{1}{2}\mathrm{log}_{2}\rho$ at high SNR. In
addition, for both Fig. 4 and Fig. 5, the advantage is the
greatest for the case $\alpha_{SR}>\alpha_{RD}$.

\begin{figure}
\begin{center}
\includegraphics [width=90mm,height=75mm]{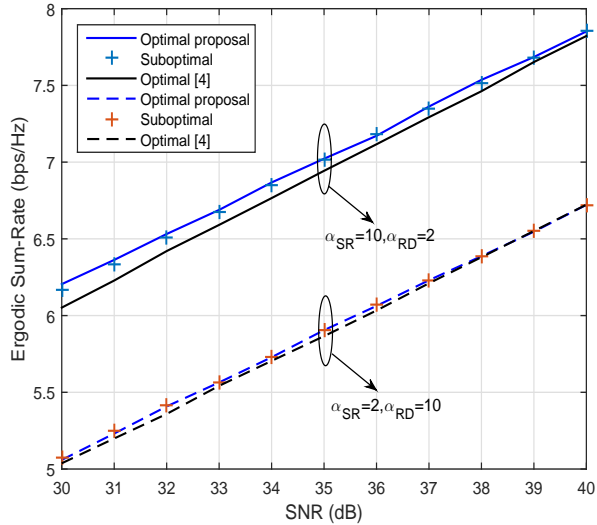}\\
\caption{\label{9} The ergodic SRs achieved by our proposed scheme
and CRS-NOMA \cite{1} with the optimal and the suboptimal power
allocation schemes versus the high transmit SNR.}
\end{center}
\end{figure}


\section{Conclusions}

In this paper, we have proposed a two-stage power allocation CRS
using NOMA and derived the closed-form solution for the proposed
system. By means of simulation results, it has been shown that our
proposed scheme significantly improves the ergodic SR compared
with the CRS-NOMA. In addition, a suboptimal solution obtained by
the practical power allocation scheme has been also provided,
which is close to the optimal one at high SNR.

\end{spacing}
\end{document}